\documentstyle[aps,preprint,epsf]{revtex}

\begin{document}
\draft

\preprint{Cond-mat/00;;;;}
\title{\bf
    Negative Magneto-Resistance Beyond Weak Localization in
    Three-Dimensional Billiards: Effect of Arnold Diffusion
}
\author{Jun Ma and Katsuhiro Nakamura}
\address{
Department of Applied Physics, Osaka City University, Sumiyoshi-ku,
Osaka 558-8585, Japan}
\date{\today}
\maketitle
\begin{abstract}
We investigate a semiclassical conductance for ballistic open
three-dimensional (3-d) billiards. For partially or completely broken-ergodic
3-d billiards such as SO(2) symmetric billiards, the dependence of the 
conductance on
the Fermi wavenumber is dramatically changed by the lead  orientation.
Application of a symmetry-breaking weak magnetic field brings about mixed
phase-space structures of 3-d billiards which ensures a novel Arnold diffusion
that cannot be seen in 2-d billiards. In contrast to the 2-d case, the 
anomalous
increment of the conductance should inevitably include a contribution arising
from Arnold diffusion as well as a weak localization correction.
Discussions are devoted to the physical condition for observing this 
phenomenon.

\end{abstract}

\narrowtext

\section{Introduction} \label{sec1}

Concave and convex billiards as prototypes of conservative chaotic systems
have received a growing theoretical and experimental interest in the fields of
nonlinear dynamics and statistical mechanics [1].
These  billiards are nowadays fabricated as quantum dots or antidots in
ballistic microstructures where  the system size is much less than
the  mean free path $\ell(\sim{20}\mu$m) and
larger than the Fermi wavelength ($\lambda\sim{50}$nm)[2].
Since the striking experiments on the magneto-conductance of nanoscale
stadium and circle billiards with a pair of conducting leads [2],
quantum transport phenomena have been intensively investigated [3]. Some
salient
features of experiments well agreed with the semiclassical theory, where a
semiclassical conductance formula was derived for two-dimensional
(2-d) chaotic billiards by combining the semiclassical Green function with
the Landauer formula [4,5]. Subsequently, the semiclassical conductance formula
was extended for integrable 2-d billiards [6], and detailed numerical
computation on circle billiards
indicated that the conductance also depends on the  lead orientation
[7].  Recently, the dependence of the conductance on
the  lead orientation  was  found experimentally in a ballistic square
billiard [8].
 
At present, another important object of research on quantum transport
is the three-dimensional (3-d) billiard [9].
The ballistic 3-d quantum dots will be
fabricated, for instance, by exploiting drying etching processess
with focused-ion beams applied to $\rm{Al}_x\rm{Ga}_{1-x}$
$\rm{As}/\rm{GaAs}/\rm{Al}_x\rm{Ga}_{1-x}\rm{As}$ double
hetero-structures, which will be less than both the elastic mean
free path and phase coherence length. The subject of this paper is to extend
the semiclassical conductance
formula for 2-d billiards  to 3-d billiards and to investigate the quantum
transport in open 3-d billiards from a viewpoint of the Arnold diffusion, a key
concept in high-dimensional nonlinear dynamical systems. While
Arnold diffusion is a classical phenomenon, an
anomalous phenomenon like this in classical mechanics should have a
quantum counterpart, which can only be
analytically revealed by using the semiclassical theory.
 
\section{Semiclassical conductance for open 3-dimensional billiards}

The  conductance for open 3-d billiards connected to a pair of rectangular
parallel-piped lead wires
can be expressed by the transmission coefficient $T$ as
$G=\frac{2e^2}{h}T=\frac{2e^2}{h}\sum_{{\bf n,m}=1}|t_{\bf nm}|^2$, where
$t_{\bf nm}$ is the {\bf S}-matrix element between the incoming mode
${\bf m}=(m_x,m_y)$
and the outgoing mode ${\bf n}=(n_x,n_y)$, and the double summation
is taken over all propagating modes. The
transmission amplitude connecting those modes at the Fermi energy $E_F$
reads as [10,11]
\begin{equation}
t_{\bf {nm}}=-i\hbar\sqrt{v_{\bf {n}}v_{\bf {m}}}
{\int}dx^{''}dy^{''}dx^{'}dy^{'}\psi^*_{\bf {n}}(x^{''},y^{''})
G(x^{''},y^{''},z^{''};x^{'},y^{'},z^{'};E_F)\psi_{\bf {m}}(x^{'},y^{'}),
\end{equation}
where $v_{\bf {m}}$ and $v_{\bf {n}}$ are the  longitudinal velocities of
electrons
for the incoming and outgoing modes, respectively.
    $x^{'},y^{'},z^{'}$ and $x^{''},y^{''},z^{''}$ are the $local$
coordinates for
     the    transverse  ($x,y$) and longitudinal ($z$) directions
     (inward the billiard for $z^{'}$ and outward for  $z^{''}$)
     of the incoming
     and   outgoing leads, respectively,
     and $G(x^{''},y^{''},z^{''};x^{'},y^{'},z^{'};E_F)$
is  the Green function for an electron propagating  from the entrance to
the exit.
$\psi_{\bf {m}}(x^{'},y^{'})$ and $\psi_{\bf {n}}(x^{''},y^{''})$  are
the transverse components of the
wave functions at the leads. To simplify the   problem, the cross section of
the leads is assumed to be a
square with side length $l$. Then   $\psi_{\bf m}(x^{'},y^{'})=
\frac{2}{l}\sin(m_x{\pi}x^{'}/l)\sin(m_y{\pi}y^{'}/l)$ and
$\psi_{\bf n}(x^{''},y^{''})=
\frac{2}{l}\sin(n_x{\pi}x^{''}/l)\sin(n_x{\pi}y^{''}/l)$. In the ballistic
regime,
the Green function from entrance (${\bf r}^{'}$) to exit
(${\bf r}^{''}$)  can be
replaced by the Gutzwiller semiclassical path-integral expression [12]
\begin{equation}
G(x^{''},y^{''},z^{''};x^{'},y^{'},z^{'};E_F)=\frac{2\pi}{(2\pi{i}{\hbar})^2
v_F}\sum_{s({\bf r}^{'},{\bf r}^{''})}|\det(\frac{\partial{\bf p^{'}_{\perp}}}
{\partial{\bf r^{''}_{\perp}}})
|^{1/2}\exp(\frac{i}{\hbar}S_s({\bf r}^{''},{\bf r}^{'},E_F)-i\pi\nu_s/2),
\end{equation}
     where $S_s$ and $\nu_s$ are the action and Maslov index for a trajectory
$s$,  and   $v_F$ is the Fermi velocity.
    The above semiclassical Green function was widely applied to derive the 
level
    density  of both chaotic and regular  billiards [12,13].
    Here, to  avoid direct transmission without any bouncing at the billiard
wall we may place a stopper inside the billiard.

    In the integration in Eq.(1) we introduce the entrance
    angle  $\theta^{'}$
     between the initial direction of the orbit  and the $z^{'}$ axis, and the
angle $\varphi^{'}$ between
    the projection of the  initial direction  onto the $x^{'}-y^{'}$ plane
and the $x^{'}$ axis. Angles
$\theta^{''}$ and $\varphi^{''}$ are  defined for the exit in the same way.
     Then,  we can rewrite $|\det(\frac{\partial{\bf p^{'}_{\perp}}}
{\partial{\bf r^{''}_{\perp}}})|$ for each trajectory $s$
     as
    $|\det(\frac{\partial{ p^{1'}_{\perp}},{ p^{2'}_{\perp}})}
    {\partial({ r^{1''}_{\perp}},{ r^{2''}_{\perp}})})|=p_F^2A_s$ where
    $A_s=|\frac{\sin\theta^{'}}{\cos\theta^{''}}(\frac{\partial\varphi^{'}}
    {\partial{x}^{''}}\frac{\partial\theta^{'}}
    {\partial{y}^{''}}-\frac{\partial\varphi^{'}}
    {\partial{y}^{''}}\frac{\partial\theta^{'}}
    {\partial{x}^{''}})|$. For large mode numbers, the integration over the
    transverse coordinates in Eq.(1) can be  performed in the stationary-phase
approximation,
and the transmission
     amplitude becomes
\begin{equation}
t_{\bf {nm}}=-i\hbar\frac{16\pi}{l^2p_F}\sum_{s({\bar m}_x,{\bar m}_y,
{\bar n}_x,{\bar n}_y)}
{\rm sgn}({\bar m}_x){\rm sgn}({\bar m}_y){\rm sgn}({ {\bar n}_x})
{\rm sgn}({{\bar n}_y})(\cos\theta^{'}\cos\theta^{''})^{1/2}
({A_s}/{B_s}) ^{1/2}e^{\frac{i}{\hbar}{\tilde S}_s-i{\tilde \nu}_s\pi/2}
\end {equation}   with
\begin {eqnarray}
B_s&=&|[\sin2\varphi^{''}
(\frac{\partial\theta^{''}}{\partial{y}^{''}}\frac{\partial\theta^{''}}
{\partial{x}^{''}}\cos^2\theta^{''}-
\frac{\partial\varphi^{''}}{\partial{y}^{''}}\frac{\partial\varphi^{''}}
{\partial{x}^{''}}\sin^2\theta^{''})+\sin2\theta^{''}(
\frac{\partial\varphi^{''}}{\partial{y}^{''}}\frac{\partial\theta^{''}}
{\partial{x}^{''}}\cos^2\varphi^{''}-
\frac{\partial\theta^{''}}{\partial{y}^{''}}\frac{\partial\varphi^{''}}
{\partial{x}^{''}}\sin^2\varphi^{''})]\nonumber\\
&\times&[\sin2\varphi^{'}
(\frac{\partial\theta^{'}}{\partial{y}^{'}}\frac{\partial\theta^{'}}
{\partial{x}^{'}}\cos^2
\theta^{'}-
\frac{\partial\varphi^{'}}{\partial{y}^{'}}\frac{\partial\varphi^{'}}
{\partial{x}^{'}}\sin^2\theta^{'})
+\sin2\theta^{'}
(\frac{\partial\varphi^{'}}{\partial{y}^{'}}\frac{\partial\theta^{'}}
{\partial{x}^{'}}\cos^2\varphi^{'}-
\frac{\partial\theta^{'}}{\partial{y}^{'}}\frac{\partial\varphi^{'}}
{\partial{x}^{'}}\sin^2\varphi^{'})]|,
\end{eqnarray}
where ${\bar m}_x=\pm{m}_{x}$,${\bar m}_y=\pm{m}_{y}$,
     ${\bar n}_x=\pm{n}_{x}$,${\bar n}_y=\pm{n}_{y}$, and
$\tilde{S}_s=S({\bf r^{''}}_s,{\bf r^{'}}_s,E_F)+\hbar\pi{{\bar m}_x}x^{'}_s/l+
\hbar\pi{{\bar m}_y}y^{'}_s/l-\hbar\pi{{\bar n}_x}x^{''}_s/l-
\hbar\pi{{\bar n}_y}y^{''}_s/l$, and
$\tilde{\nu}$ is the Maslov index  which includes an   extra
phase coming from the possible sign change in each of the four Fresnel
integrals.
The above result means that only those isolated trajectories connecting
a pair of transverse planes at discrete angles
$\sin\theta^{'}\cos\varphi^{'}={\bar m}_x\pi/k_Fl,
\sin\theta^{'}\sin\varphi^{'}={\bar m}_y\pi/k_Fl,
\sin\theta^{''}\cos\varphi^{''}={\bar n}_x\pi/k_Fl$, and
$\sin\theta^{''}\sin\varphi^{''}={\bar n}_y\pi/k_Fl$  dominate the conductance.

Using Eqs.(3,4), we shall first analyze a completely chaotic 3-d billiard with
a pair of conducting leads.
In this case, the  electron injected from the incoming  lead will bounce
at the billiard boundary in an ergodic way before reaching the exit,
leading  to amplitudes of the same order
for both transmission and reflection coefficients.
In the large mode number case, the summations over modes are replaced
by integrations and the transmission coefficient
$T=\sum_{\bf {n,m}}|t_{\bf {nm}}|^2$
is rewritten as
\begin{eqnarray}
T&=&{\hbar^2}\frac{256\pi^2}{l^4}
\frac{1}{p_F^2}\sum_{{\bar m}_x,{\bar m}_y,{\bar n}_x,{\bar n}_y}\sum_{s,u}
\cos\theta^{'}\cos\theta^{''}(\frac{A_sA_u}{B_sB_u})^{1/2}e^{\frac{i}{\hbar}
(\tilde{S}_s-\tilde{S}_u)-i\frac{\pi}{2}
(\tilde{\nu}_s-\tilde{\nu}_u)}\nonumber\\
&\approx&\frac{256\pi^2}{l^4k_F^2}
\int\int_0^{({\bar n}_x^2+{\bar n}_y^2)\leq\frac{k_F^2\l^2}{\pi^2}}{d}
{\bar n}_x{d}{\bar n}_y
\int\int_0^{({\bar m}_x^2+{\bar m}_y^2)\leq\frac{k_F^2\l^2}{\pi^2}}{d}
{\bar m}_x{d}{\bar m}_y
\sum_{s,u}\cos\theta^{'}\cos\theta^{''}\nonumber\\
&\times&(\frac{A_sA_u}{B_sB_u})^{\frac{1}{2}}
e^{\frac{i}{\hbar}(\tilde{S}_s-\tilde{S}_u)-i\frac{\pi}{2}
(\tilde{\nu}_s-\tilde{\nu}_u)}\nonumber\\
&=&\frac{256{k_F^2}}{\pi^2}\int^{1}_{-1}\sin\theta^{'}d(\sin\theta^{'})
\int^{\pi}_{-\pi}d\varphi^{'}\int^{1}_{-1}\sin\theta^{''}d(\sin\theta^{''})
\int^{\pi}_{-\pi}d\varphi^{''}\sum_{s,u}\cos\theta^{'}\cos\theta^{''}
\nonumber\\
&\times&(\frac{A_sA_u}{B_sB_u})^{\frac{1}{2}}e^{\frac{i}{\hbar}
(\tilde{S}_s-\tilde{S}_u)-i\frac{\pi}{2}
(\tilde{\nu}_s-\tilde{\nu}_u)}.
\end{eqnarray}
Noting the dimension of $(\frac{A_sA_u}{B_sB_u})^{1/2}$ to be $l^2$, and using
the diagonal approximation, the above integration can be
evaluated as $T\propto(k_Fl)^2$. The conductance for 3-d chaotic billiards
is thus given
by $G\propto\frac{e^2}{h}(k_Fl)^2$ without any dependence on the lead
orientations.

\section{Completely or partially broken-ergodic 3-d billiards}

However, the situation will be dramatically changed for the case of
completely or partially broken-ergodic 3-d billiards. As examples, we choose
SO(2)-symmetric billiards: A completely integrable (: broken-ergodic)
3-d billiard is available by rotating, e.g., the 2-d ellipse
billiard (on the $Y-Z$ plane) around the $Z$ axis (see Fig.1). Note:
$X,Y,Z$ are
the $global$ coordinates; A partially  broken-ergodic 3-d billiard is also
available in a similar way  by rotating, e.g., the 2-d stadium. All these
SO(2)-symmetric 3-d
billiards have the angular momentum $L_Z$ as a constant of motion.
In an open system version of each of these billiards, one may place in its
inside
a smaller stopper of the similar shape, to avoid direct transmission of
electrons. The resultant 3-d $shell$ billiards still retain the SO(2) symmetry.
 
       Firstly, consider the vertical case (i) when the incoming and outgoing
leads are connected
with the billiard vertically at points $A$ and $C$, respectively (see Fig.1).
Electrons to reach the exit $C$ should have the vanishing angular momentum,
$L_Z=0$.
Therefore, only the electrons with the initial velocity vector at the
entrance $A$
lying in the $Y-Z$ plane can reach $C$  because of the conservation of $L_Z$;
Other incoming electrons falling into  the trajectories out of this plane
cannot reach $C$ and should return to $A$.  In deriving the transmission
coefficient
for this lead orientation, the integration prior to Eq.(5) is reduced to
the line
integration 
\begin{figure}[tbh]
\begin{center}
\noindent
\leavevmode\epsfxsize=120mm
\epsfbox{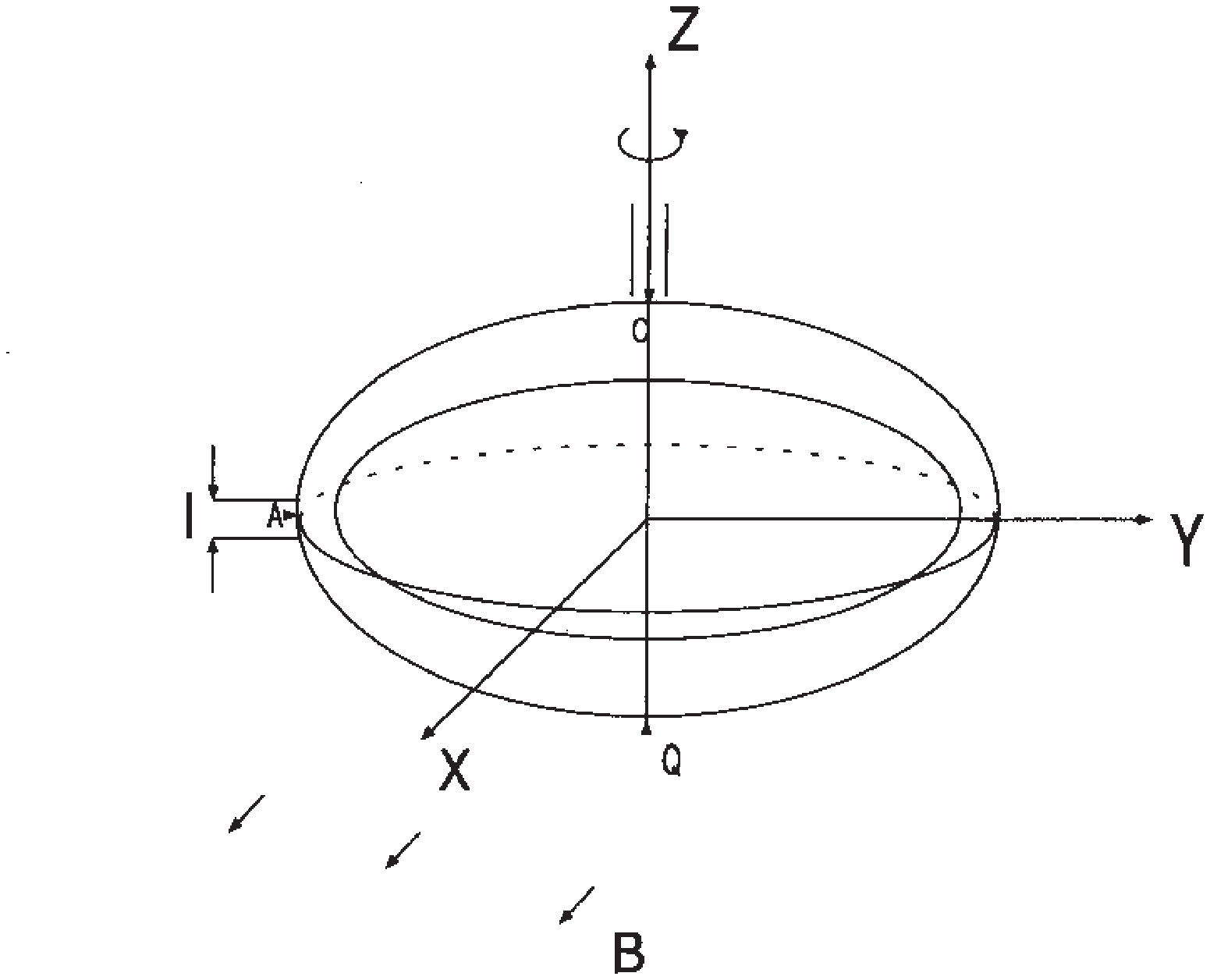}
\caption{Schematic illustration of the 3-d elliptic shell billiard.
Dual
    2-d ellipses sharing the same foci in $Y-Z$ plane are
    rotated around the $Z$ axis to
generate a 3-d  elliptic shell billiard.
The longer  radii of outer and inner ellipses are taken as $2\Re$ and
$1.85\Re$.
    The shorter radius of the outer ellipse is $0.8\Re$. The side length $l$
of the square
lead is taken as $\Re/20$.
}
\end{center}
\end{figure}
\begin{flushleft}over $\theta^{'}$ and $\theta^{''}$ only, yielding the conductance
$G\propto\frac{e^2}{h}{k_Fl}$ (linear in $k_F$!). (The reflection 
\end{flushleft} coefficient,
which is obtained by applying the same procedure
as in the chaotic billiard, is proportional to $({k_Fl})^2$.)
We shall then consider  the parallel case (ii) when the incoming and outgoing
leads are parallelly placed at points $Q$ and $C$, respectively (see Fig.1).
Each electron incoming from the entrance $Q$ has the vanishing angular
momentum,
$L_Z=0$, and thereby can reach $C$ or return to $Q$ with probability of
the same order. Therefore, for this lead orientation,
$G\propto\frac{e^2}{h}({k_Fl})^2$.

Generally, in completely or partially  broken-ergodic 3-d billiards, we have
$G\propto\frac{e^2}{h}(k_Fl)^\gamma$ with
$0<\gamma\le2$, and the exponent $\gamma$ is determined by how
the billiard symmetry is broken by the lead orientation.
It should be noted: open 2-d billiards always share
$G\propto\frac{e^2}{h}{k_F}l$ [4], independently of both integrability and
lead orientations.

\section{Effects of symmetry-breaking weak magnetic field}

The magneto-resistance is also a very important problem in 3-d billiards.
It is well known that in ballistic 2-d billiards the decrease of resistance
with increasing magnetic field (: negative magneto-resistance) occurs due to
the weak localization based on the quantum interference between a pair of
time-reversal symmetric orbits[2-5]. What kind of additional novel
phenomena will be
expected in SO(2)-symmetric 3-d  billiards with the vertical lead ($A,C$)
orientation when a symmetry-breaking weak magnetic field $B$ will be applied
along the $X$ axis?  In contrast to phase-space structures in 2-d systems where
each of chaotic zones generated by weak $B$ field is mutually separated by the
Kolmogorov-Arnold-Moser (KAM) tori, those in 3-d systems consist of
chaotic zones mutually connected
via narrow chaotic channels called the Arnold web (AW) [14-16]. In this case,
electrons incoming in an arbitrary direction, which permeate (chaotic) zone
by zone in an ergodic way through the AW, can show a global diffusion, i.e.,
Arnold diffusion (AD). Therefore, a $B$ field increases the transmission
channels and the number of stationary points in the integration in Eq.(3),
and the resultant integration in Eq.(5) over $\varphi^{'}$ and $\varphi^{''}$
as well as $\theta^{'}$ and $\theta^{''}$ indicates
$G\propto\frac{e^2}{h}({k_Fl})^2$.
In general we shall expect a more general result, $G\propto(k_Fl)^\beta$ with
$1<\beta\le2$, and the exponent $\beta$ depends on the width of AW.
A crucial point is that a $B$ field has increased the
exponent in the power-law behavior of the semiclassical conductance in case
of the vertical lead orientation. This anomalous phenomenon is just the
negative magneto-resistance beyond the weak localization. We here keep to 
employ
the terminology of the $negative$ $magneto-resistance$ wherever a
$B$ field reduces the zero-field resistance.

On the other hand, for the parallel lead ($Q,C$) orientation where each 
electronic
trajectory lies almost in the vertical plane, $G\propto\frac{e^2}{h}(k_Fl)^2$
is retained, irrespective of the absence or presence of $B$ field; We can 
expect
only the $B$ field-induced increment of the proportionality constant, which
implies a conventional negative magneto-resistance caused by suppression of the
weak localization.

Before moving to the next section, we should note:
Originally, the negative magneto-resistance was evidenced in the numerical
analysis of the conductance of 2-d quantum dots [4]. The accompanying 
theory, based
on the resistance, manipulated backscattering orbits and attributed this 
phenomenon to
the weak localization [4]. Later works [5] based on the conductance, which 
introduced
the small-angle-induced diffraction effect together with
an idea of interference between a pair of $partially$ time-reversal symmetric
orbits [5], swept away a cloud (i.e., breakdown of unitarity) hanging
over the above theory. All our results in the present section have been 
obtained
just within the same primitive analysis of the conductance as the original one
for 2-d dots.

\section{Numerical results for open 3-d elliptic billiard}
 
To numerically verify the above prediction, we shall choose a SO(2)-symmetric
3-d elliptic shell billiard with the vertical lead orientation.
This ellipsoid is a typical example of completely integrable SO(2)-symmetric
3-d billards. The system in the presence of a symmetry-breaking
field $B$ parallel to $X$ axis will be examined. To understand
the classical dynamics of electrons,
we analyze as a Poincar\'e surface of section the $X-Z$ plane
with longitude $S$ and velocity component $v_Z/v^0$ as Birkhoff coordinates
[17]. ($v^0=\frac{\hbar\pi}{lm_e}$ is the velocity unit corresponding to the
lowest mode.) For $B=0$, the trajectory from the entrance $A$ with an 
arbitrary initial
velocity vector lying out of $Z-Y$ plane proves to be confined to a torus 
(see Fig.2(a)),
failing to reach the exit $C$, i.e., $S=\pi/2$.  Let $\eta$ ($\equiv\Re/\Re_c$)
be the ratio between the  cyclotron radius $\Re_c$ and the characteristic
length $\Re$ of the cross-sectional ellipse (see the caption of Fig.1).
Then $B$ field can be expressed as
$B=\eta{B}_0$ with $B_0=m_ev^0/e\Re$ ($B_0$ corresponds to $\Re$). In case of
a weak field ($B=0.02B_0$), Poincar\'e sections
for the 
\begin{figure}[tbh]
\begin{center}
\noindent
\leavevmode\epsfxsize=100mm
\epsfbox{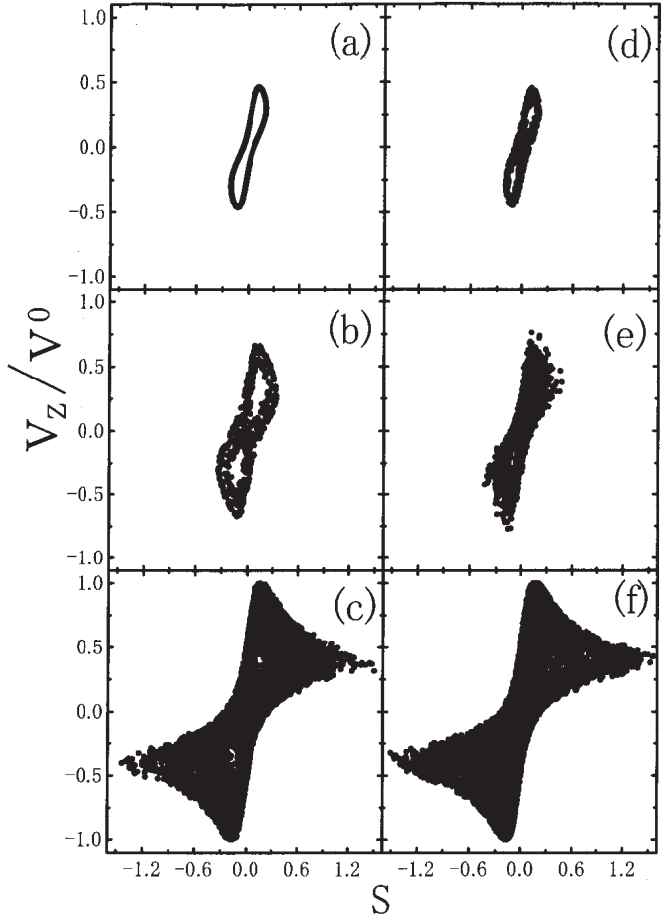}
\caption{Poincar\'e sections ($X-Z$ plane)
     for the trajectories emanating from point $A$.
       Longitude $S$ ($-\pi/2\le{S}\le\pi/2$) and velocity component $v_Z/v^0$
    ($v^0=\frac{\hbar}{lm_e}$) are chosen as Birkhoff coordinates. $S=0$ and 
$S=\pi/2$
   imply entrance $A$ and exit $C$, respectively.
    (a) $B=0$ and initial velocity $v_X/v^0=0.821,
    v_Y/v^0=0.549, v_Z/v^0=0.157$; (b),(c) $B=0.02B_0$ and the same initial
    velocity as (a): (b)$P=500$; (c)$P=17000$.
     (d)-(f)  $B=0.02B_0$ and initial velocity
    $v_X/v^0=0.843, v_Y/v^0=0.538, v_Z/v^0=0$: (d)$P=500$; (e)$P=4100$;
    (f)$P=16000$. $P$ is the number  of bouncings at the outer billiard
    intersecting $X-Z$ plane. For the meaning of $B_0$, see the text.
}
\end{center}
\end{figure}
\begin{flushleft}trajectory with the same initial velocity vector as in 
Fig.2(a)
are given in Figs.2(b)  and (c) \end{flushleft}
for increasing numbers of bouncings at
the outer billiard wall, and similarly those for the trajectory with the
initial velocity vector lying in the $X-Y$ plane are given in Figs.2(d)-(f).
One can find:  as the $B$ field is switched on, the orbit trajectory
first leaves the initial torus,  enters an outer layer, and,
repeating a similar process, diffuses over more and more distant layers,
exhibiting a phenomenon of the Arnold diffusion (AD).
Even though a symmetry-breaking field is weak enough to keep
the orbit almost straight, AD makes it possible for the trajectories with
initial angular momenta $L_Z\ne0$
to reach the exit $C$, i.e., $S=\pi/2$ (see Figs.2(c) and (f)) and eventually
to contribute to conductance.

\begin{figure}[tbh]
\begin{center}
\noindent
\leavevmode\epsfxsize=130mm
\epsfbox{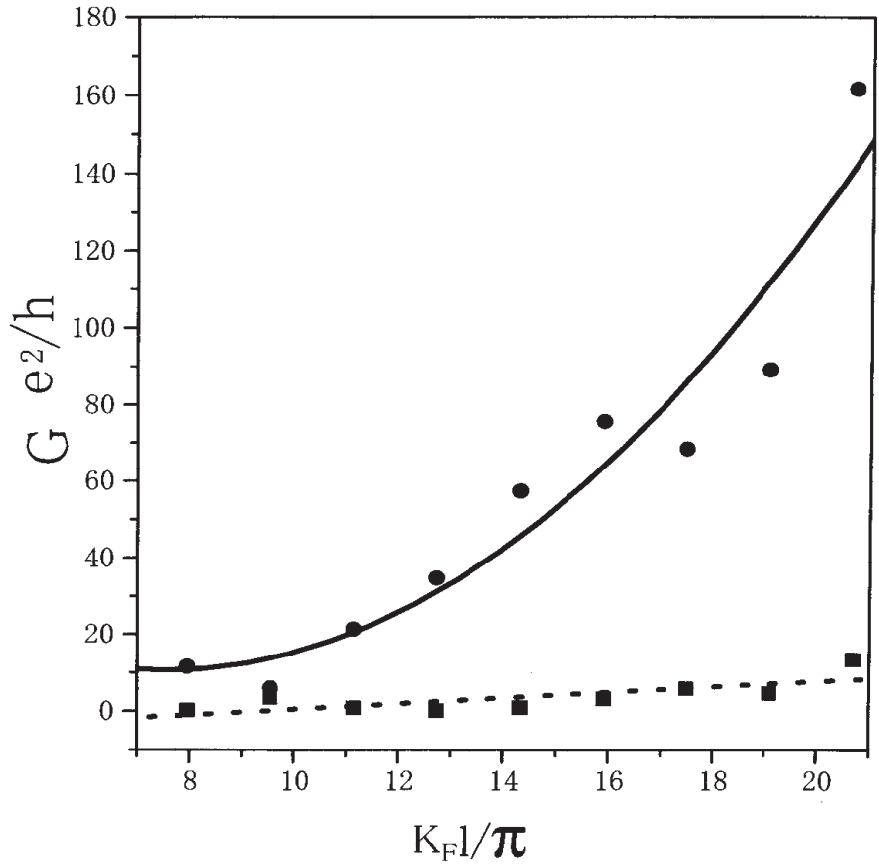}
\caption{Numerical results for semiclassical conductance versus Fermi
wavenumber
    for $A,C$ leads orientation. Circle and square symbols  are
     for $B=0.02B_0$ and $B=0$, respectively. Solid quadratic and dashed linear
      lines are the corresponding fitted curves for their means.}
\end{center}
\end{figure}

      To proceed to the numerical calculation of semiclassical transmission
amplitudes
in Eqs.(3,4) for the 3-d elliptic billiard with the vertical lead orientation,
let us define the $4\times4$  monodromy matrix
${\bf {M}}$ as
$\left ( \begin{array}{c}
{\delta {\bf r}_{\perp}^{''}} \\
{\delta {\bf p}_{\perp}^{''}}
\end{array}
\right )
={\bf {M}}
\left ( \begin{array}{c}
{\delta {\bf r}_{\perp}^{'}} \\
{\delta {\bf p}_{\perp}^{'}}
\end{array}
\right )$.
At the entrance we introduce
a pair of mutually orthogonal coordinates ${{ r}^1_{\perp}}^{'}$ and
${{ r}^2_{\perp}}^{'}$ lying in
the transverse plane perpendicular to
the initial direction of each orbit. Similarly
${{ r}^1_{\perp}}^{''}$ and ${{ r}^2_{\perp}}^{''}$ are chosen
at the exit. Using elements (${{M}}_{ij}$) of ${\bf {M}}$ matrix,
we rewrite the necessary factors for
the numerical computation. For example,
$\frac{\partial\theta^{''}}{\partial{x^{''}}}
=\frac{\partial(\theta^{''},y^{''})}
{\partial({r^1_{\perp}}^{'},{r^2_{\perp}}^{'})}
\frac{\partial({r^1_{\perp}}^{'},{r^2_{\perp}}^{'})}
{\partial({r^1_{\perp}}^{''},{r^2_{\perp}}^{''})}
\frac{\partial({r^1_{\perp}}^{''},{r^2_{\perp}}^{''})}
{\partial(x^{''},y^{''})}=\frac{1}{p_F}[{{M}}_{41}(
\cos\varphi^{''}\cos\theta^{''}{{M}}_{12}-\sin\varphi^{''}{{M}}_{22})-
{{M}}_{42}(
\cos\varphi^{''}\cos\theta^{''}{{M}}_{11}-\sin\varphi^{''}{{M}}_{21})]/({{M}
}_{11}{{M}}_{22}-{{M}}_{12}{{M}}_{21})$ and
$\frac{\partial\theta^{'}}{\partial{x}^{''}}=
\frac{\partial(\theta^{'},y^{''})}{\partial({r^1_{\perp}}^{''},{r^2_{\perp}}
^{''})}
\frac{\partial({r^1_{\perp}}^{''},{r^2_{\perp}}^{''})}
{\partial(x^{''},y^{''})}=\frac{\cos\theta^{''}}{p_F}(\sec\varphi^{''}
\csc\theta^{''}/{{M}}_{14}+\csc\varphi^{''}/{{M}}_{24})$.
While ${{\bf M}}$ and $\nu$ (Maslov index) can be computed by the method
suggested
in Refs.[9,18], the first and last rotation angles in ${\bf {M}}$
should be determined by using ${{ r}^1_{\perp}}^{'}$ , ${{ r}^2_{\perp}}^{'}$
and ${{ r}^1_{\perp}}^{''}$ , ${{ r}^2_{\perp}}^{''}$
chosen as  above, since, after folding the orbit, the
initial and final coordinates should become identical.
Combining all these factors with Eqs.(3,4), the semiclassical conductance is
calculated explicitly. The result, which includes both the
classical contribution and the quantum correction in an inseparable way, is
plotted in Fig.3. Although the data show small fluctuations, the fitted curves
for their means are described by $G\propto\frac{e^2}{h}(k_Fl)^2$
and $G\propto\frac{e^2}{h}{k_Fl}$ in cases of $B=0.02B_0$ and $B=0$, 
respectively,
confirming the analytic issue suggested in the previous section.

The increase of the exponent in the $k_F$ dependence
of $G$ upon switching on a weak field, which cannot be seen in 2-d cases,
provides a numerical evidence that
the negative magneto-resistance (: a difference between two lines in Fig.3)
in 3-d systems comes from AD as well as the weak-localization correction.

\section{Summary and discussions}

We have investigated semiclassical quantum transport in three-dimensional
(3-d) ballistic quantum billiards, and explored novel phenomena that cannot 
be seen
in 2-d billiards. For partially or completely broken-ergodic 3-d billiards
such as SO(2) symmetric  billiards, the dependence of the conductance $G$ on
the Fermi wavenumber $k_F$ is dramatically changed by the lead orientation.
In case of the vertical orientation,
a symmetry-breaking weak magnetic field turns out to increase the exponent
in $k_F$ dependence of $G$.
In marked contrast to the 2-d case, the anomalous increase of the
conductance should be caused by the Arnold diffusion (AD)
as well as a weak localization correction.

We should note the following:(i)In an extremely weak field case when
the width of Arnold web (AW) is thin enough,
we shall see $G\propto(k_Fl)^\beta$ with $1<\beta<2$. Our numerical 
calculation in
the previous section suggests that, to have $\beta=2$ in a very large $k_F$
regime, $\Re/\Re_c$ should be no less than $10^{-2}$; (ii) role of a stopper is
also essential.
Reflection at an inner convex wall should increase the instability of orbits,
which is favourable for a genesis of AD; (iii) to clearly observe AD,
the symmetry-breaking field should destroy all constants of motion except
for the energy. From this viewpoint,
it is more advantageous to choose a SO(2)-symmetric 3-d chaotic shell billiard
generated from a fully-chaotic 2-d billiard with a hollow by its rotation
around the $Z$ axis. The billiards of this kind, which are partially 
broken-ergodic,
are experimentally more accessible than a (completely broken-ergodic) 
elliptic shell
billiard, and all conclusions for the latter should hold for the former as 
well.
 
Finally, we point out the role of temperature in observing
the effect of AD on the conductance. At finite temperatures,
the conductance is related to the scattering matrix as
$G=\frac{2e^2}{h}\int^{\infty}_{-\infty}
d\epsilon[-\frac{{d}f}{d{\epsilon}}]\sum_{\bf n,m}|t_{\bf n,m}|^2$
with $f$ the Fermi distribution function. Then the integration over energy
will give  a temperature factor
$\frac{\pi{k}_{\rm B}TL/(\hbar{v}_{\rm F})}
{\sinh(\pi{k}_{\rm B}TL/(\hbar{v}_{\rm F}))}$
with $L$ the length of trajectories. For the vertical
lead orientation, this factor suppresses very long trajectories reaching
the outgoing lead along the Arnold web, and thereby the negative 
magneto-resistance
is described only by the weak-localization correction. On the other hand, 
if the
temperature is lowered, the self-averaging over $k_F$ is not permissible,
leading to a suppression of the weak-localization correction.
Then the increment of the magneto-resistance is mainly caused by the effect
corresponding to the Arnold diffusion. We hope that a future experiment
on 3-d ballistic cavities can verify our prediction.

\end{document}